\documentclass[twoside]{ilcws08}
\usepackage[latin1]{inputenc}
\usepackage[dvips]{graphicx,epsfig,color}
\usepackage{wrapfig,rotating}
\usepackage{amssymb,amsmath,array}

\begin{document}
\title{
Improvements to the ILC Upstream Polarimeter} 
\author{Jenny List and Daniela K\"afer
\thanks{The authors acknowledge the support of the DFG under LI 1560/1-1.}
\vspace{.3cm}\\
DESY - FLC \\
Notkestr. 85, 22607 Hamburg - Germany
}

\maketitle

\begin{abstract}
The physics programme of the ILC requires polarimetry with yet unprecedented precision.
This note focusses on aspects of the upstream polarimeter as described in the ILC Reference Design Report which are not compatible with the extraordinary precision goals. In conclusion, recommendations for improving the design are given. 
\end{abstract}

\section{ILC polarimetry}

Since the ILC is designed to allow measurements of masses and cross-sections of Standard Model as well as of possible new particles at the permille level, also the beam parameters like beam energy, polarisation and luminosity have to be controlled to this precision. 
While for the beam energy this goal has already been achieved at previous colliders, the up to now most precise polarisation measurement at SLD reached a precision of 0.5\%~\cite{LEPSLD}. 
The overall polarimetry scheme at the ILC therefore combines the measurements of two Compton polarimeters, located upstream and downstream of the $e^+e^-$ interaction point, with data from the $e^+e^-$ annihilations themselves. For optimized Compton polarimeters, a factor of two improvement over the SLD is expected. 

\section{The upstream polarimeter}

\begin{wrapfigure}{r}{0.6\columnwidth}
\centerline{\includegraphics[width=0.6\columnwidth]{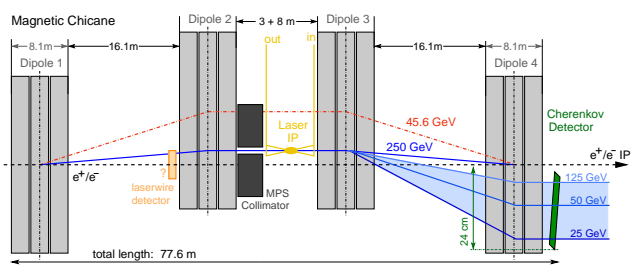}}
\caption{The upstream polarimeter.}\label{Fig:chicane}
\end{wrapfigure}

The working principle of Compton polarimeters has been described in detail for example in~\cite{Tesla}. 
The lonigtudinally polarised electron (or positron) beam is hit under a small angle by a circularly polarised laser. The energy spectrum of the scattered particles depends on the product of laser and beam polarisations. The rate asymmetry with respect to the laser helicity is directly proportional to the polarisation: The analyzing power, which contains all dependency on the experimental setup, corresponds to the asymmetry which is expected for 100\% polarisation. Obviously a large analyzing power is favourable for a precise measurement.

In order to collect statistics fast enough, the laser intensity is chosen such that typically in the order of 1000 electrons are scattered per bunch. Since the scattering angle in the laboratory frame is restricted to less than 10~$\mu$rad, a magnetic chicane has to be employed to transform the energy spectrum into a spacial distribution, which is finally measured with an array of Cherenkov detectors. 
Figure~\ref{Fig:chicane} shows the setup forseen for the ILC upstream polarimeter, with the Compton IP between the second and third dipole triplet. The blue line shows the beam trajectory for a beam energy of 250~GeV, the blue shade indicates the fan of (detectable) Compton scattered electrons. 
 
Based on the experience from SLD, it has to be expected that the largest contribution to the overall error budget is due to the analyzing power. However with the dedicated chicane design forseen at the ILC and with a detector operating without a preradiator, this uncertainty is expected to be reduced by about a factor 2 with respect to SLD, yielding a contribution of 0.2\%. To achieve this goal it is important to controle all system parameters continuously. For example a misalignment of the polarimeter detector with respect to the beam of 0.5~mm leads to an 0.1\% effect on the polarisation. Apart from the analyzing power, additional sources of uncertainty are non-linearities of the detector and its read-out chain (0.1\%) and the measurement of the laser polarisation (0.1\%).

\section{Scaled versus fixed field operation}

\begin{wrapfigure}{r}{0.7\columnwidth}
\vspace{-0.75cm}
\centerline{\includegraphics[width=0.35\columnwidth]{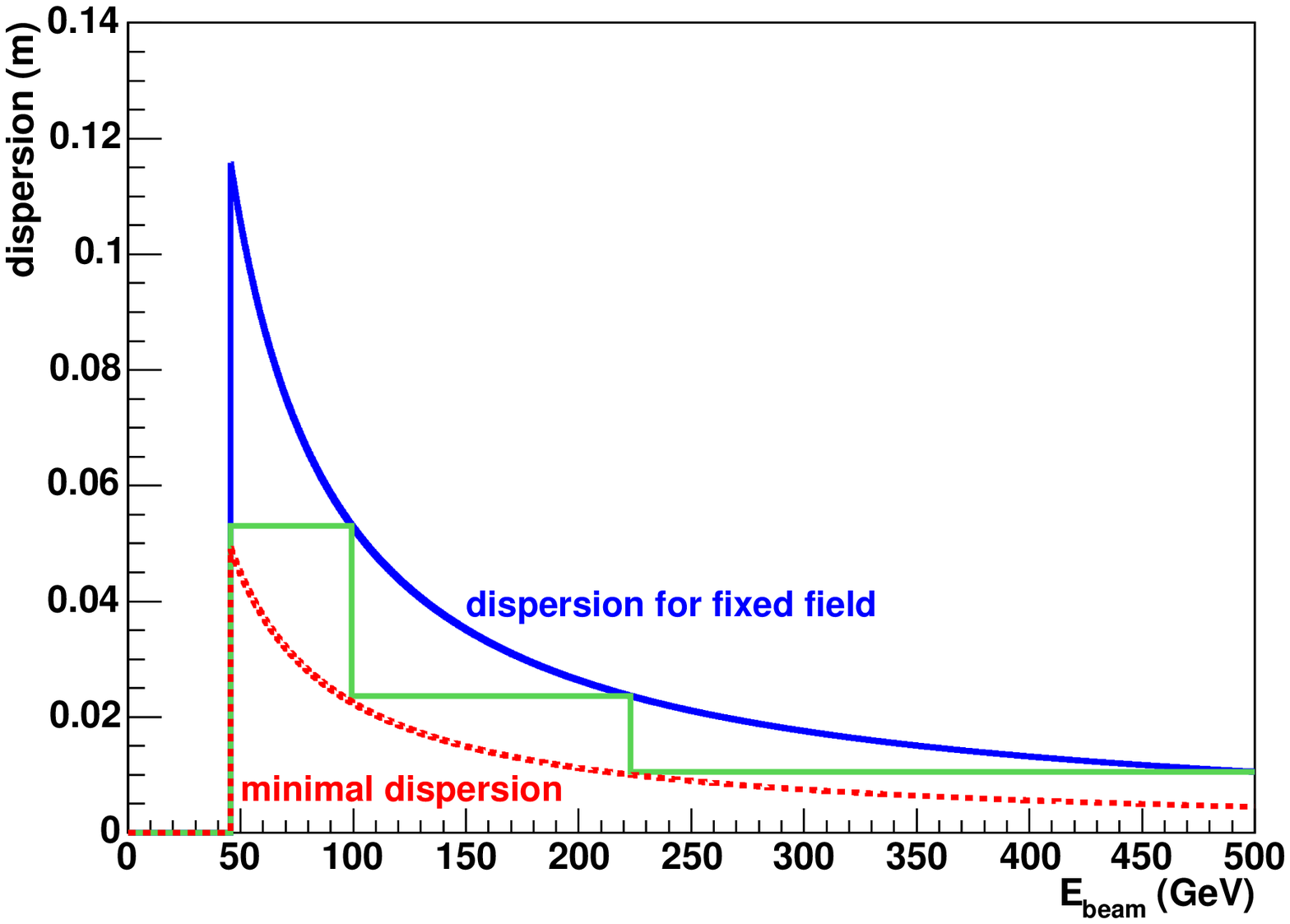}
\hspace{-0.05cm}
\includegraphics[width=0.35\columnwidth]{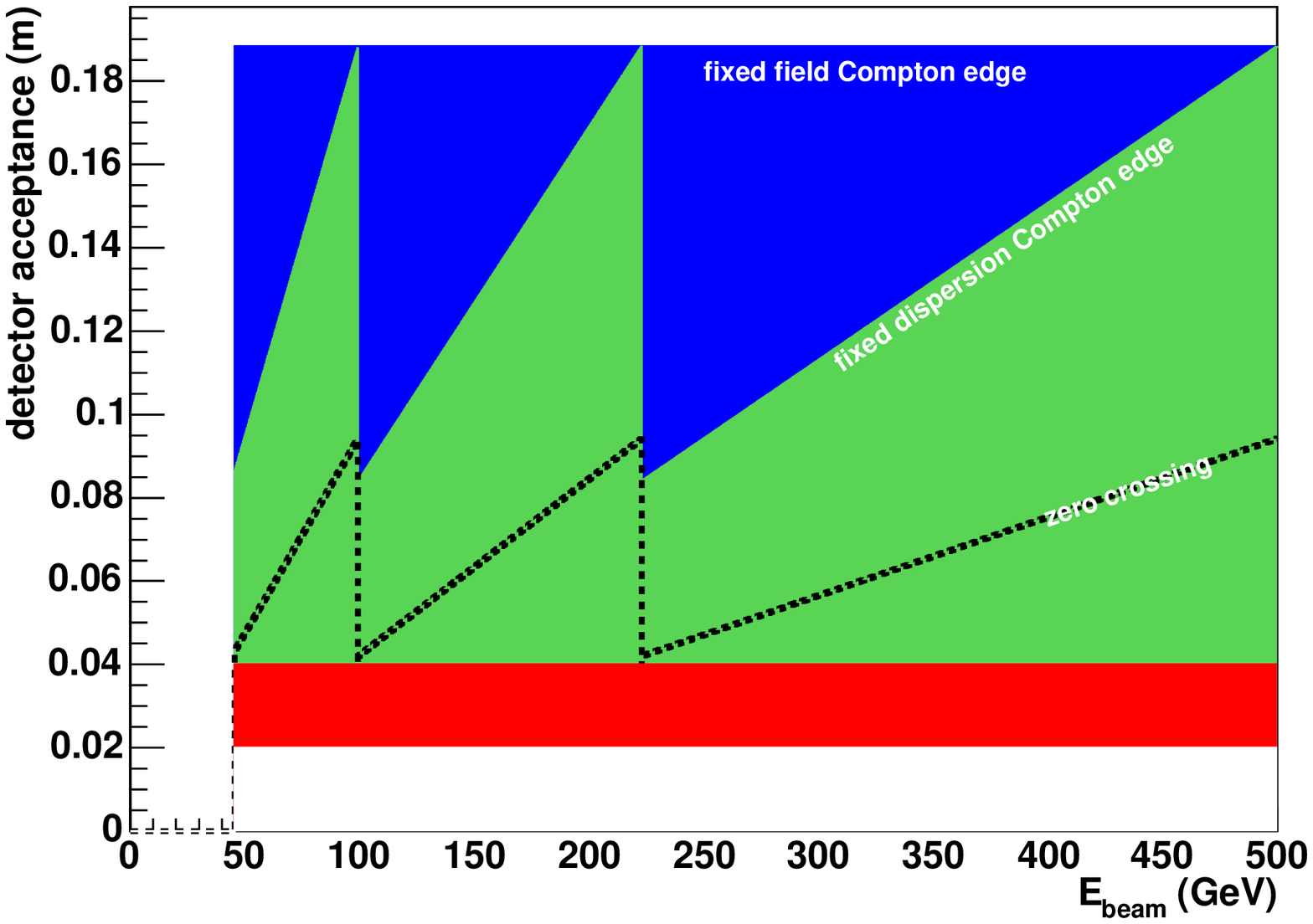}}
\caption{Dispersion of the chicane (left) and detector coverage (right) as a function of the beam energy.}\label{Fig:disp_cov}\end{wrapfigure}

If operated at a fixed magnetic field for all beam energies, the position of the Compton edge at the detector surface stays the same for all beam energies, which ensures a homogeneous detector acceptance and the same measurement quality at all center-of-mass energies. Instead, the Compton IP moves laterally with the beam energy, as indicated in Figure~\ref{Fig:chicane} by the red dashed line for 45.6~GeV. While the laser path can be easily adjusted with movable mirrors, the MPS collimator, would be more difficult and expensive to move. Therefore it has been proposed to operate the chicane with a magnetic field which scales with the beam energy, thus keeping the Compton IP at a constant location. In such a scenario, the Compton edge position varies with the beam energy, squeezing the entire Compton spectrum more and more into the beampipe until no measurement is possible anymore. 

 Figure~\ref{Fig:disp_cov} shows the dispersion of the chicane, i.e. the transverse distance of the Comtpon IP from the neutral beam line, as  function of the beam energy for the original fixed field case (blue line). It has been chosen to maximize the spread of the Compton spectrum at the detector position while avoiding a significant emittance blow-up. The red dashed line indicates the minimum dispersion required to perform any polarisation measurement at all (two detector channels below zero crossing of asymmetry). With fixed dispersion, the whole range of ILC beam energies from 45.6~GeV up to 500~GeV can only be covered by at least three ranges, indicated by the green line. The minimal scale factor for the magnetic field with respect to the fixed field case is 0.45. 

In addition, Figure~\ref{Fig:disp_cov} shows the detector coverage as a function of the beam energy. In the fixed field scenario, the Compton spectrum is spread out over the whole colored area, where the red area correponds to the two innermost detector channels, while the dashed line indicates the zero crossing of the asymmetry. In the three step scaled field scenario, the Compton spectrum would only cover the red and green areas. In this case, the achievable precision will depend on the beam energy.

The analyzing power of each detector channel depends on the Compton edge position with respect to the beam. For a precise polarisation measurement, it is therefore of utmost importance to controle this parameter at any time, preferably without interfering with polarimetry data taking. The fewer detector channels contribute to the measurement, the more sensitive the polarisation measurement becomes to the Compton edge position. 

\begin{wrapfigure}{l}{0.6\columnwidth}
\vspace{-0.75cm}
\centerline{\includegraphics[width=0.3\columnwidth]{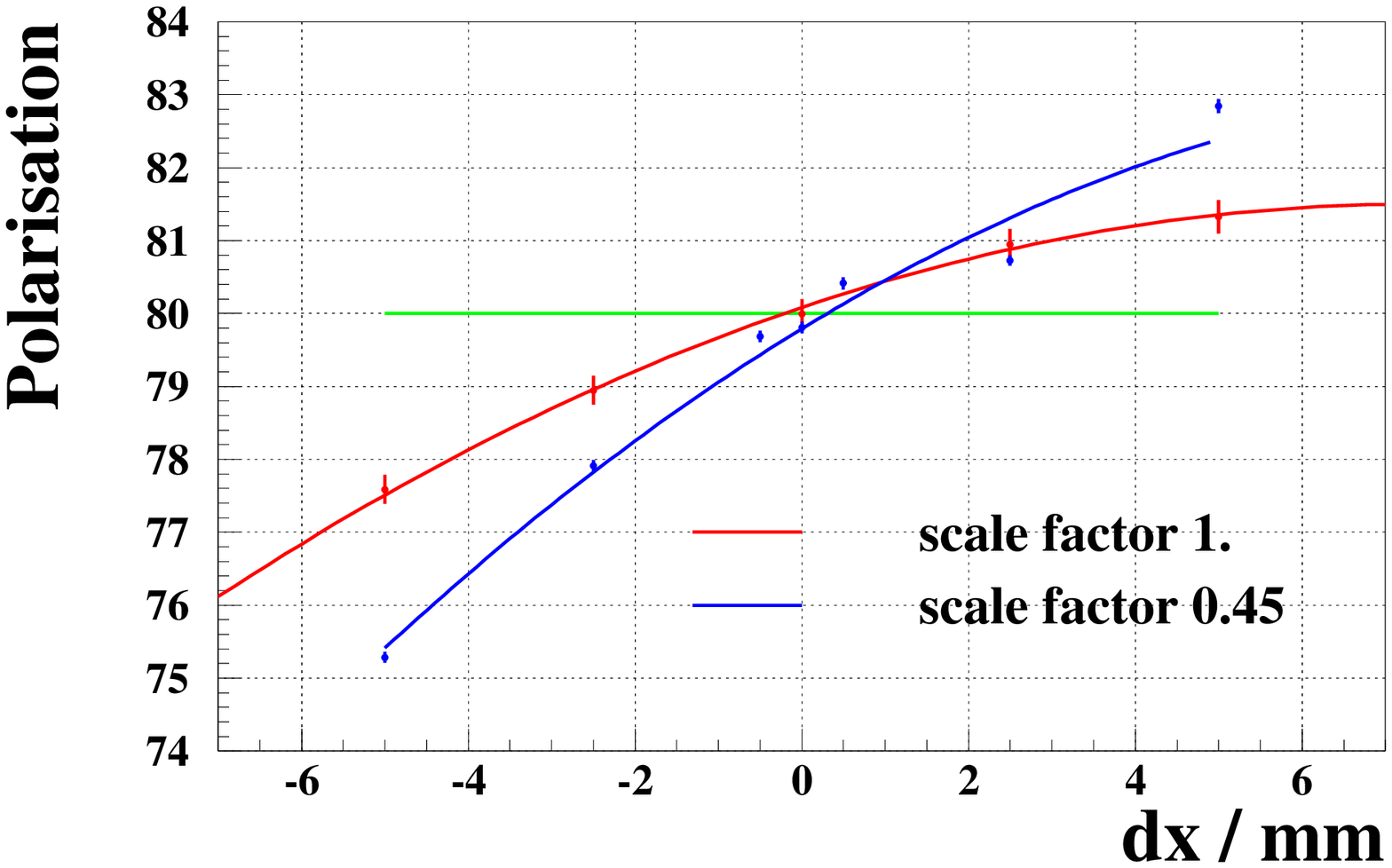}
\hspace{-0.05cm}
\includegraphics[width=0.3\columnwidth]{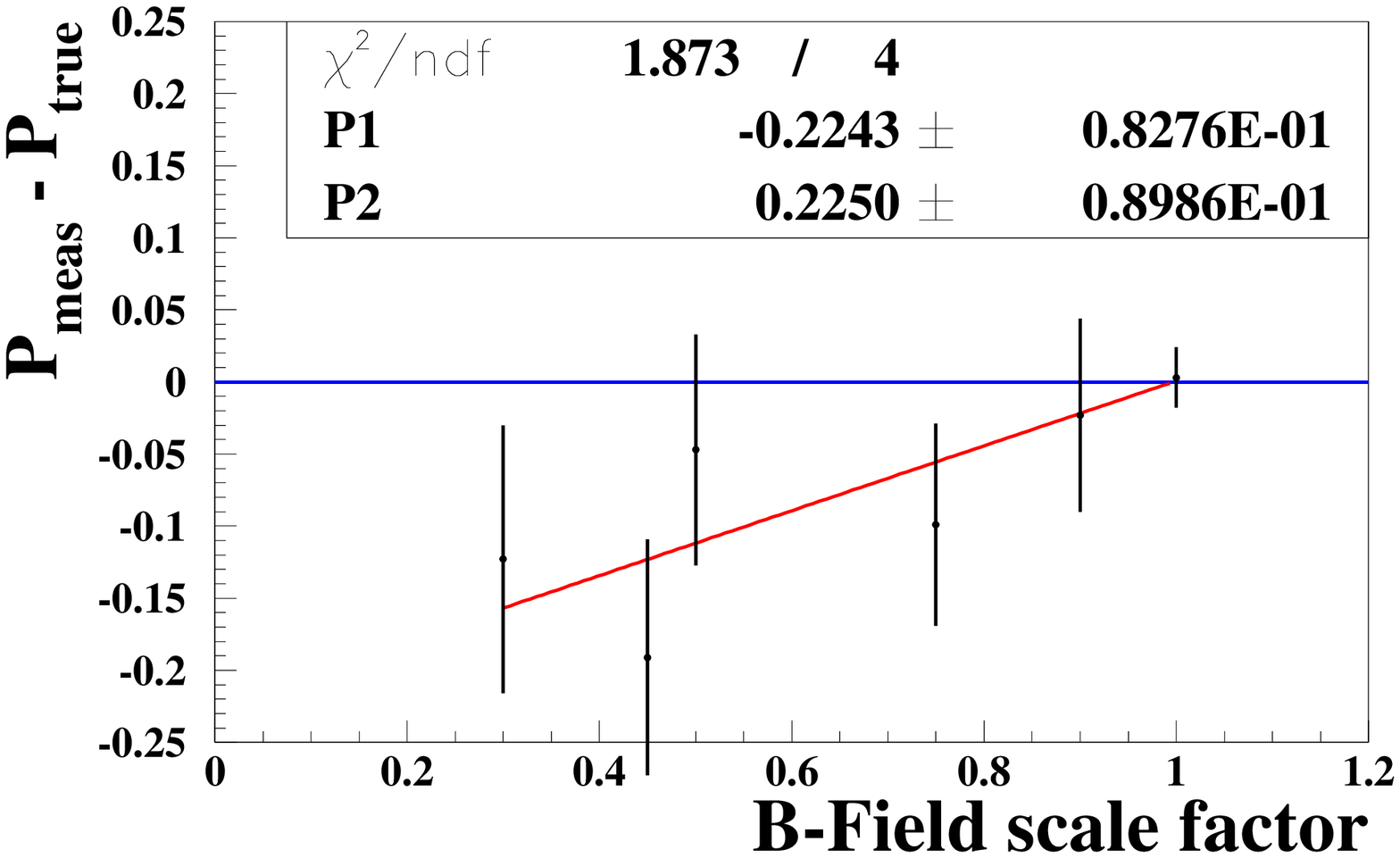}}
\caption{Left: Reconstructed polarisation as function of a miscalibration of the Compton edge's position for two different scale factors. Right: Difference between input and reconstructed polarisation as function of the magnetic field scale factor.}\label{Fig:edge_calib}\label{Fig:pol_bias}\end{wrapfigure}

This is illustrated in figure~\ref{Fig:edge_calib}, which shows the effect of some simulated misalignment $dx$ between detector and beam for an input polarisation of 80\% (green line). The red curve corresponds to the case of a fixed magnetic field, where the Compton spectrum is spread out over about 20~cm. If the edge position should not contribute more than 0.1\% to the total error budget, it needs to be known to better than 0.4~mm. In the scaled field scenario as discussed above, the magnetic field is reduced down to 0.45 times the nominal value. In this case (blue line), the dependence of the polarisation on the alignment is much steeper, since each channel integrates a larger fraction of the highly non-linear spectrum, and the uncertainty on the edge position must not exceed 0.2~mm in order to stay within less than 0.1\% deviation of the polarisation.

On the other hand, the effective position resolution of the detector gets worse if less channels are covered by the spectrum. With a simple algorithm which estimates the position of the edge within the last covered bin by compairing its contents with the expectation extrapolated from its neighboring bins, the edge position can be determined to $x_{edge} = (19.760 \pm  0.024 (stat) \pm 0.003 (syst))$~cm in the fixed field case after 10000 bunch crossings, where the difference to the true position is given as systematic uncertainty. In the scaled field case, again at a scale factor of 0.45, the same method yields $x_{edge} = (8.622 \pm  0.010 (stat) \pm 0.271 (syst))$~cm, showing clearly that the difference to the true edge position is large enough to introduce a bias of more than 0.1\% on the polarisation measurement. This is illustrated by Figure~\ref{Fig:pol_bias}, which shows the difference between reconstructed and input polarisation as a function of the scale factor, using the edge position determined from the simulated data instead of its true value.
 
\section{Conflicts with collimator and emittance diagnostics}

\begin{wrapfigure}{r}{0.51\columnwidth}
\centerline{\includegraphics[width=0.525\columnwidth]{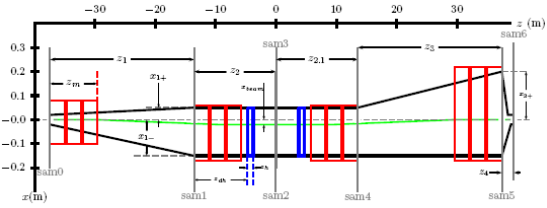}}
\caption{Tapered vaccuum chamber as designed for the fixed field scenario.}\label{Fig:vac_chamber}
\end{wrapfigure}

Figure~\ref{Fig:vac_chamber} shows a sketch of the vaccuum chamber forseen for the fixed field scenario. At the last dipole triplet, it has to provide an aperture of about 30~cm for the fan of scattered Compton electrons as well as for the trough going beam. In order to avoid wake fields, the aperture grows slowly along the entire chicane. The effect of a collimator at a place where the aperture of the vacuum chamber is already about 20~cm is currently unclear. The collimator will create significant backgrounds for the polarimeter detector, which is not compliant with a high precision measurement. Last but not least, the collimator either would have to be movable or a scaled field operating mode of the chicane is required, putting in danger precision polarimetry, as explained in the previous section.

In addition to the collimator it has been suggested to install laser wire based emittance diagnostics right upstream of the polarimeter chicane, with a detector for the scattered particles in front of the second dipole triplet, c.f. Figure~\ref{Fig:chicane}. This detector is expected to create backgrounds at the level of 60\% of the polarimeter signal~\cite{lawrence}. Such amounts of background are clearly incompatible with meaningful polarimetry. Thus, the laser wire and polarimeter could only be operated on alternating bunches. Alternating operation however cannot resolve the spacial conflicts: If the Compton scattered photons from the laser wire are used, a converter target plus an electron detector is needed in order to distinguish the Compton photons from the synchrotron background coming out of the linac. At high beam energies, the dispersion of the chicane is only a few centimeters (c.f. Figure~\ref{Fig:disp_cov}), yielding not enough clearance for converter and detector with respect to the through-going beam. The obvious alternative is to detect directly the Compton electrons, which are deflected out of the neutral beam line by the first dipole triplet of the chicane. This approach seems promissing, especially since it doesn't create additional backgrounds for the polarimeter. However, if the chicane is operated with a scaled field, there is not enough beam clearance at low beam energies. In summary, the best compatibility of emittance diagnostics and polarimetry in one chicane is achieved when operating on alternating bunches, with a fixed magnetic field and electron detection for the laser wire.

\section{Conclusions}

The upstream polarimeter as it is forseen in the ILC RDR has several short comings, which make it impossible to reach the high precision goals. Especially the chicane magnets should be operated at a constant magnetic field for all beam energies. Furthermore significant additional background sources must be avoided. Therefore it is recommended to separate the locations of the polarimeter,  the MPS collimator and the emittance diagnostics. A full list of recommendations can be found in~\cite{exsum}.





\begin{footnotesize}

\end{footnotesize}


\end{document}